\documentclass{jps-cp}
\usepackage{txfonts} 

\title{Theory on Hadrons in Nuclear Medium}

\author{Su Houng \textsc{Lee}$^{1}$}

\inst{$^{1}$Department of Physics, Yonsei University, Seoul 03722, Korea}

\email{suhoung@yonsei.ac.kr}

\recdate{January 31, 2019}

\abst{
After decades-long  attempts  to measure the mass shift and understand the origin of hadron mass,  it became clear that one has to analyze hadrons with small vacuum width.  Also, to identify the effect of chiral symmetry breaking, one has to start by looking at chiral partners.  In this talk, I will review why such consideration inevitably led us to consider $K^*$ and $K_1$ in nuclear matter\cite{Song:2018plu}.  With the kaon beam at JPARC, one could observe the mass shift of both  particles in a nuclear target experiment.  Once the masses and mass difference of $K^*$ and $K_1$ mesons are measured, we will be  closer to understanding the origin of the hadron masses and the effects of chiral symmetry breaking in them. }

\kword{Hadron mass, strange vector meson, strange axial vector meson, JPARC, Kaon beam.}

\begin{document}
\maketitle

\section{Introduction}

 The mass of a hadron can not be understood from the traditional picture of a composite particle, the mass of which can be expressed in terms of the masses of its constituents and a small binding energy.  This is because the origin of a hadron mass is intricately related to the low energy property of QCD such as  confinement and chiral symmetry breaking effects, which are analytically still not completely understood.  
It is known that chiral symmetry breaking is partly responsible for the masses \cite{Hatsuda:1985eb,Brown:1991kk,Hatsuda:1991ez,Klingl:1997kf}.  Therefore, if one can identify the effect of chiral symmetry breaking to the masses of hadron, we will be able to build model and then a theory to understand the masses of the hadrons.  

In the original in-medium QCD sum rules\cite{Hatsuda:1991ez},
the changes of the vector meson masses were dominantly due to the change of the  four-quark condensate for the light quark system and the strange quark condensate for the strange quark system.
Experiments have since then been performed worldwide to observe mass shift of hadrons at finite temperature or density \cite{Hayano:2008vn}. The first hint of vector meson mass shift was first observe by the KEK-PS experiments from analyzing the invariant mass spectra of $e^+~e^-$ pairs from the nuclear targets.  The experiment found excess signals at the lower end of the $\omega$ resonance peak that could be  explained by the vector meson  mass decrease of 9 $\%$ at normal nuclear density~\cite{Naruki:2005kd}. The possible mass shift of the $\phi$ meson has also been searched for in experiments. 
In fact, evidence of mass shift by 3.4 $\%$
at normal nuclear density was reported by the  KEK-PS E325 collaboration~\cite{Muto:2005za}.  Further measurements with higher statistics are planned at J-PARC by the  E16 experiment~\cite{Ichikawa:2018woh}.

Dilepton spectrum do not suffering from strong final state interaction with the medium when the signal emanates   from inside the nucleus but is low in the yield and has many sources. Reactions involving hadronic decays  have the opposite features.
By analyzing the reaction $\gamma+A \rightarrow \omega +X \rightarrow \pi^0 \gamma +X'$, the CBELSA/TAPS Collaboration observed that the $\omega$ meson mass decreased by about 60 MeV
 at an estimated average nuclear density of 0.6 $\rho_0$~\cite{Trnka:2005ey}.
 However,  the final state hadronic interaction could have contaminated the signal.   Also, the QCD sum rule approach was found to have a large contribution from the scattering term \cite{DuttMazumder:2000ys,Thomas:2005dc,Leupold:2009kz},  making the sum rule analysis less reliable.
In the mean time, a successful set of experiments were performed by  the CBELSA/TAPS collaboration  to extract meson-nucleus optical  potentials
 from near-threshold meson productions reactions.  The success comes  from focussing on  mesons with small width in the vacuum ($K, \eta, \eta', \omega, \phi$)~\cite{Metag:2017yuh}.  Furthermore, the problem with final state interaction was found to be solvable by combining the measurement of excitation functions and the  transparency 
ratios.

Motivated by such recent experimental progresses, the author has  recently estimated the spectral shift of the $f_1$ meson, which is a chiral partner of $\omega$  in the limit where the  disconnected diagrams are neglected, in the QCD sum rule approach~\cite{Gubler:2016djf}.
Experimentally,   the $f_1(1285)$ has been successfully identified by the CLAS collaboration in photoproduction
from a proton target with a small width of $18 \pm1.4$ MeV \cite{Dickson:2016gwc},  so that
 performing similar experiments on a nuclear target and comparing the result with that from the  $\omega$  would be extremely useful for
  partial restoration of chiral symmetry in nuclei as suggested in Ref.~\cite{Gubler:2016djf}.
 In fact,  the individual meson mass could behave differently depending on whether the hadron  is in nuclear matter or at finite temperature,
 but the mass difference between chiral partners will only depend on the chiral order parameter and should be universal~\cite{Lee:2013es}.

However, while $\langle \bar{q}q \rangle$ is the order parameter of QCD,  it is not clear how to directly relate them to the masses of the hadrons.  
On the other hand, one can show that the difference between the vector and axial-vector correlation functions in the open strange channel is also an order parameter of chiral symmetry~\cite{Lee:2013es}.  This implies that their spectral densities will become degenerate if chiral symmetry is restored.  In the vacuum, the low-lying modes that couple to the vector current are  $K^*(892)$  and $K^*(1410)$ while for the axial vector current they are  $K_1(1270)$ and  $K_1(1400)$. There is a subtlety in the nature of the two $K_1$ states: They are assumed to be a mixture of the $^3P_1$ and $^1P_1$ quark-antiquark pair in the quark model \cite{Suzuki:1993yc}. However, if chiral symmetry is partially restored, the spectral density will tend to become degenerate so that the lowest distinctive poles in the respective current will approach each other. Therefore, in a recent work~\cite{Song:2018plu}, we investigated the spectral modification of open strange meson $K_1$ through the axial-vector current in nuclear matter using QCD sum rules.

Measuring the  open strange meson in the vector channel, namely the $K^{*+}$  through the  decay  $K^{*+} \rightarrow K^+ + \gamma$ was suggested as a promising signal to measure the spectral change of the vector meson in Ref.~\cite{Hatsuda:1997ev}.
Both the $K_1(1270) $ and the $K^{*}(892)$ have widths smaller than their non strange counter parts, namely 90 MeV and 47 MeV, respectively, compared to more than 250 MeV and 150 MeV for the $a_1$ and the $\rho$. At the same time, they are also chiral partners so that their mass difference is sensitive to the chiral order parameter.

We note that  $K_1^+$ and $K_1^-$ become non-degenerate in nuclear medium due to the  presence of nucleons which break charge conjugation  invariance in the medium. This is not a problem as long as we compare $K^*$ and $K^1$ with the same charge states as the chiral partnership exits between them.

\section{Importance of condensate}

Let us first discuss the importance of condensate in discussing the hadron masses.  We are still far from understanding the low energy physics directly from QCD variables.   Yet, important physical effects can be summarized in terms of local operators.  The most important operators are chiral order parameters: that is operators that characterizes the breaking of chiral symmetry.  There are many chiral order parameters.  Below are some of the most commonly cited order parameters. 
\begin{eqnarray}
\langle \bar{q} q \rangle,~~ \langle( \bar{q} \gamma_\mu \tau^a q)^2-(\bar{q} i \gamma_5 \gamma_\mu \tau^a q)^2 \rangle, \cdot \cdot \cdot.
\end{eqnarray}
The method of constructing chiral order parameters are discussed in the next section.  

On the other hand, there are no local operators that are directly related to  confinement.  Confinement is related to the Wilson  loops. In particular, the area-law behavior in pure gauge theory can be related to the confinement physics.  It is well known that while the space-time Wilson loops changes from the area law behaviour to the perimeter law behaviour as one increased the temperature across the  the deconfinement temperature $T_c$, the area-law behaviour of space-space Wilson loop does not change\cite{Manousakis:1986jh}.  There were several attempts to related the area-law behaviour to gluon condensates\cite{Shifman:1980ui,Dosch:1988ha}.  In Ref. \cite{Lee:1989qj}, the author has therefore studied the non-perturbative part of the gluon condensate across  $T_c$ and found that while the condensate values changes, about 50\% of the vacuum value survives above $T_c$, suggesting that non-perturbative nature survives as was seen in the persistent area-law behaviour of the space-space Wilson loop.  However, when the total condensate values including the perturbative contributions are extracted\cite{Lee:2008xp}, one finds that while the electric part of the gluon condensate values changes abruptly across $T_c$, the magnetic part does not change much.  This behavior seems quite in line with what is seen in the space-time (space-space) Wilson loop where the leading condensate in their respective operator production expansion (OPE) is the $\langle \frac{\alpha_s}{\pi} E^2 \rangle$   ($\langle \frac{\alpha_s}{\pi} B^2 \rangle$).

\section{Chiral order parameter}

The commonly known chiral order parameter can be rewritten in several forms.
\begin{eqnarray}
\langle \bar{q} q \rangle & \equiv & \lim_{x \rightarrow 0} 
- \langle {\rm Tr} [ S(0,x) ] ] \rangle =   \lim_{x \rightarrow 0} 
- \frac{1}{2} \langle {\rm Tr} [ S(0,x)- i\gamma_5  S(0,x) i \gamma_5 ] ] \rangle  \nonumber \\ 
&  = & -\pi \langle \rho(\lambda=0) \rangle, 
\label{BC-0}
\end{eqnarray}
where the second line shows the density of zero eigenvalue in the Euclidean formalism known as the Banks-Casher formula\cite{BC}.   The gauge invariant part of the combination of propagators 
$(S(x,y)-i\gamma^5 S(x,y) i  \gamma^5)$ can be identified as a 
convenient expression of a chiral order parameter.  
The second line, due to Banks-Casher, is useful to identify the origin or chiral symmetry breaking in terms of quark eigenvalues that can be usefully implemented in a lattice gauge theory calculations.

Using Eq.~(\ref{BC-0}) one can identify many more order parameters of chiral symmetry breaking.  A set of order parameters can be obtained by looking at the difference in the correlation functions of chiral partners.   
For example the difference between the scalar and pseudo scalar correlation function is an  order parameter\cite{Cohen:1996ng}.
\begin{eqnarray}
\Delta_{S-P} (q) & = & \int d^4x e^{iqx} \langle \mathcal{T} \bigg( \bar{q} \tau^a q(x) \bar{q} \tau^a q(0)- \bar{q} \tau^a i\gamma^5 q(x)  \bar{q} \tau^a i\gamma^5 q(0) \bigg) \rangle  \nonumber \\
& = & \int d^4x  e^{iqx} \frac{1}{2} \langle {\rm Tr}  [ \tau^a \tau^a \bigg( i \gamma^5 S(x,0) i \gamma^5 -S(x,0) \bigg) \bigg( S(0,x)-  i \gamma^5 S(0,x) i \gamma^5  \bigg) ] \rangle  .  \label{p-s-0}
\end{eqnarray}
As can be seen above, the combination of propagators appear, which will be proportional to the density of zero modes.  Using these methods, one can construct chiral order parameters of different types of correlation functions.
We have put in the flavor matrix in the scalar current.  In principle, this is not necessary because the additional disconnected diagram in the scalar channel is also proportional to the chiral order parameter. 
\begin{eqnarray}
\Delta_{S}^{disconnected} (q) & = & \int d^4x e^{iqx} \langle \mathcal{T} \bigg( \bar{q} q(x) \bar{q}  q(0)\bigg) \rangle_{disconnected}  \nonumber \\
& = & \int d^4x  e^{iqx} \frac{1}{4} \langle {\rm Tr}  [ \bigg( i \gamma^5 S(x,x) i \gamma^5 -S(x,x) \bigg) ] {\rm Tr} [\bigg( S(0,0)-  i \gamma^5 S(0,0) i \gamma^5  \bigg) ] \rangle  . \nonumber \\
&& \label{p-s-1}
\end{eqnarray}

The difference between the vector and axial vector in the Kaon sector is also on order parameter.  In this case, the difference is proportional to the the chiral symmetry breaking in both flavours.  
\begin{eqnarray}
\Delta_{V-A}^m (q) & = & \int d^4x e^{iqx} \langle \mathcal{T} \bigg( \bar{s} \gamma_\mu q(x) \bar{q}  \gamma_\mu s(0)- \bar{s}  \gamma^5 \gamma_\mu  q(x)  \bar{q}  \gamma^5 \gamma_\mu s(0) \bigg) \rangle  \nonumber \\
& = & \int d^4x  e^{iqx} \frac{1}{2} \langle {\rm Tr}  [ \gamma_\mu \bigg(  i\gamma^5  S_q(x,0) i \gamma^5 -S_q(x,0) \bigg) \gamma_\mu \bigg( S_s(0,x)-i \gamma_5 S_s(0,x) i \gamma_5 S(0,x)   \bigg) ] \rangle  , \nonumber \\ &&
\label{v-a}
\end{eqnarray}
where the subscripts $q,s$ in the propagator denotes the flavour.
It should be noted that chiral partners in this case are between $K^*,K_1$ with the same flavour structure.  That is between $K^*(\bar{s}q), K_1(\bar{s}q)$ or between $K^*(\bar{q}s), K_1(\bar{q}s)$.

The vector mesons and the widths  are given in the table 1.  The $\rho$ and $a_1$ are chiral partners, but with the large width, any previous attempts to measure their mass shift seem futile.  The $\omega$ and the $f_1(1285)$ have small width, but are isospin singlets and are not chiral partners as discussed in Ref.~\cite{Gubler:2016djf}.  Still, in the limit where we neglect the flavor changing disconnected diagrams, they form chiral partners and hence experimental measurements are extremely useful as they provide input for extracting the  density dependence of condensates.  
On the other hand, $K^*,K_1$ are chiral partners but also have reasonably small widths.  Furthermore, their first excited states are already degenerate so that the chiral symmetry breaking effects seems to be concentrated in the ground states. Therefore, we will concentrate on $K^*,K_1$.
\begin{table}[tbh]
\caption{Mass and width of vector mesons. Units are MeV.}
\label{t2}
\begin{tabular}{||c|cc||c|cc||}
\hline
$J^{PC}=1^{--}$ & Mass & Width & $J^{PC}=1^{++}$ & Mass & Width \\
\hline
$\rho$ & 770 & 150 & $a_1$ & 1260 &  250 -600 \\
\hline
$\omega$ & 782 & 8.49 & $f_1$ & 1285 &  24.2 \\
\hline
$\phi$ & 1020 & 4.266 & $f_1$ & 1420 &  54.9 \\
\hline
$K^*(1^-)$ & 892 & 50.3 & $K_1(1^+)$ & 1270 &  90 \\
\hline
$K^*(1^-)$ & 1410 & 236 & $K_1(1^+)$ & 1400 &  174 \\
\hline
\end{tabular}
\end{table}

\section{$U_A(1)$ breaking effect}

There is a  cautionary point when taking the difference of correlation function.  Consider taking the difference between the scalar and pseudo-scalar  currents in SU(2) for simplicity. 
\begin{eqnarray}
\Delta_{s-\eta}^m (q) & = & \int d^4x e^{iqx} \langle \mathcal{T} \bigg( \bar{q} q(x) \bar{q}  q(0)- \bar{q} i \gamma^5   q(x)  \bar{q} i \gamma^5 q(0) \bigg) \rangle .
\label{s-eta}
\end{eqnarray}
This correlation has contribution from both the connected and disconnected  diagrams.  Also, the differences in the connected diagram is related to the chiral order parameter and does not vanish when chiral symmetry is broken.  However, even when chiral symmetry is restored, there is a contribution that is related to the topologically non-trivial configuration and does not vanish\cite{Lee:1996zy}.  These are the $U_A(1)$ breaking effects and are related to the zero modes containing all flavours.  The contribution can be well understood by looking whether the effective instanton vertex contributes\cite{tHooft:1986ooh}.  The effective vertex in the quark form is of the following form in SU(2).
\begin{eqnarray}
\det [\bar{q}_Rq_L.+h.c] = \bigg(\bar{u}_R u_L \times \bar{d}_R d_L- \bar{u}_R d_L \times \bar{d}_R u_L +.h.c. \bigg).
\end{eqnarray}
It is clear that this term interpolates between the correlation function and gives a non-vanishing contribution to Eq.~(\ref{s-eta}).  The form in the meson degrees of freedom has the following form.
\begin{eqnarray}
\det [\bar{q}_Rq_L.+h.c] \propto \bigg( \sigma^2+\pi^2-\eta^2 -\alpha^2 \bigg).
\end{eqnarray}
Hence, it is also clear that this equation will also give a  non-vanishing contribution to Eq.~(\ref{s-eta}).

\section{$K_1,K^*$ sum rules}

Let us now present the sum rule calculation for the $K^*$ and $K_1$ meson given in Ref~\cite{Song:2018plu}. The time-ordered current correlation function of the $K^*,K_1$ current is given by
\begin{eqnarray}
\Pi_{\mu \nu}(\omega, {\bf q})= i\int d^4 x e^{iq\cdot x}\langle |T[j_\mu, \bar{j}_\nu ]|\rangle,
\label{correlator}
\end{eqnarray}
where $q^\mu=(\omega, {\bf q})$ and 
\begin{eqnarray}
j_\mu^{K_1^+} =  \bar{s} \gamma_\mu \gamma_5 u &, & ~~~ j_\mu^{K_1^-}  =  \bar{u} \gamma_\mu \gamma_5 s \nonumber \\
j_\mu^{K^{*+}}  =  \bar{s} \gamma_\mu u &, &~~~ j_\mu^{K^{*-}}  =  \bar{u} \gamma_\mu s.
\label{current}
\end{eqnarray}
We will not consider currents with $d$ quarks as we will only consider symmetric matter so that the results would be the same as those with the $u$ quarks.

As the currents are not conserved, the polarization function will have contributions from the scalar and pseudo-scalar mesons.  Furthermore, there will be transverse and longitudinal polarization in the medium.  Here, we will only consider the limit where we take the external three momentum to be zero $q^\mu=(\omega,\bf{0})$.  In this work, we will extract the vector and axial-vector part by looking at the following projection. 
\begin{eqnarray}
\frac{1}{3} (q^\mu q^\nu/q^2-g^{\mu \nu} ) \Pi_{\mu \nu}(q) \stackrel{{\bf q} \rightarrow 0}{\longrightarrow} \Pi(\omega, 0). 
\label{correlator1}
\end{eqnarray}

\subsection{OPE}

\subsubsection{Vacuum}

The OPE in the vacuum has the following form up to dimension 6 operators\cite{Song:2018plu}.
\begin{eqnarray}
\Pi(q^2)=B_0 Q^2\ln \frac{Q^2}{\mu^{2} }+ B_2\ln \frac{Q^2}{\mu^2}
-\frac{B_4}{Q^2}-\frac{B_6}{Q^4},
\label{OPE}
\end{eqnarray}
where for the axial vector current,
\begin{eqnarray}
B_0 &=& \frac{1}{4\pi^2}\bigg(1+\frac{\alpha_s}{\pi}\bigg)\nonumber\\
B_2 &=& \frac{3m_s^2}{8 \pi^2}\nonumber\\
B_4^A &=& -m_s \langle\bar{u}u\rangle_0 +\frac{1}{12}\langle \frac{\alpha_s}{\pi} G^2\rangle_0 \nonumber\\
B_6^A
 &= & \frac{2 \pi \alpha_s}{9 } \bigg( \langle (\bar{s} \gamma_\mu \lambda^a s +\bar{u} \gamma_\mu \lambda^a u  )( \sum_q \bar{q} \gamma_\mu \lambda^aq )\rangle_0 \bigg) + 2 \pi \alpha_s \bigg( \langle (\bar{u} \gamma_\mu \lambda^a s )( (\bar{s} \gamma_\mu \lambda^a u ) \rangle_0
 \bigg)
 \nonumber \\
&=& \frac{32\pi \alpha_s}{81}\bigg(\langle\bar{u}u\rangle_0^2+\langle\bar{s}s\rangle_0^2\bigg)  +
\frac{32\pi \alpha_s}{9}\bigg(\langle\bar{u}u\rangle_0 \langle\bar{s}s\rangle_0\bigg). \nonumber
\end{eqnarray}
Here, the superscript $A$ in $B_n$ refers to those specific for the axial vector current. 
For the vector current $J^{K^*}_\mu=\bar{u} \gamma_\mu s$, the OPE up to this order will be similar with $B_0^V=B_0$, $B_2^V=B_2$ and
\begin{eqnarray}
B_4^V &=& +m_s \langle\bar{u}u\rangle_0 +\frac{1}{12}\langle \frac{\alpha_s}{\pi} G^2\rangle_0 \nonumber\\
B_6^V
 &= & \frac{2 \pi \alpha_s}{9 } \bigg( \langle (\bar{s} \gamma_\mu \lambda^a s +\bar{u} \gamma_\mu  \lambda^a u)( \sum_q \bar{q} \gamma_\mu \lambda^aq) \rangle_0\bigg)  + 2 \pi \alpha_s \bigg( \langle (\bar{u} \gamma_\mu \gamma^5 \lambda^a s )( (\bar{s} \gamma_\mu \gamma^5 \lambda^a u ) \rangle_0
 \bigg)
 \nonumber \\
&=& \frac{32\pi \alpha_s}{81}\bigg(\langle\bar{u}u\rangle_0^2+\langle\bar{s}s\rangle_0^2\bigg)  -
\frac{32\pi \alpha_s}{9}\bigg(\langle\bar{u}u\rangle_0\langle\bar{s}s\rangle_0\bigg). \nonumber
\end{eqnarray}
Here, $Q^2 \equiv -q^2$, $\mu=1$ GeV is the renormalization scale,
 $n$ in $B_n$ indicates the canonical dimension of the operator, and $\langle {\cal O} \rangle_0$ denotes the condensate of operator $\cal{O}$ in the vacuum. Here we take the value used recently\cite{Song:2018plu}.

\subsubsection{Medium}

In general, the correlation function in medium will have contribution from terms odd in $q^\mu$ because of the presence of the nucleons that breaks the charge conjugation invariance in the medium.  Therefore, 
\begin{eqnarray}
\Pi_1(q^2)=\Pi^e(q^2)+q_0 \Pi^o(q^2),\label{pi1eo-m}
\end{eqnarray}
where
\begin{eqnarray}
\Pi^e(q^2)&=& B_0 Q^2\ln \frac{Q^2}{\mu^{2} }+ B_2\ln \frac{Q^2}{\mu^2}
-\frac{B_4^{*V,A}}{Q^2}-\frac{B_6^{*V,A}}{Q^4},\nonumber\\
\Pi^o(q^2) &=& \frac{1}{3Q^2}(A_1^u-A_1^s)\rho
-\frac{2m_N^2}{3Q^4}(A_3^u-A_3^s)\rho, \nonumber \\
&& \label{pieo}
\end{eqnarray}
with
\begin{eqnarray}
B_4^{*A,V}& = & B_4^{A,V} +\frac{m_N}{2}(A_2^u +A_2^s)\rho \nonumber\\
B_6^{*A}& = & B_6^{A} + \frac{8 \pi}{3	q^2}q^\mu q^\nu \alpha_s \langle (\bar{u} \gamma_\mu \lambda^a s )( (\bar{s} \gamma_\nu  \lambda^a u ) \rangle_{\cal ST}-\frac{5}{6}m_N^3(A_4^u +A_4^s)\rho, \nonumber \\
B_6^{*V}& = & B_6^{V} +\frac{8 \pi}{3q^2}q^\mu q^\nu \alpha_s  \langle (\bar{u} \gamma_\mu \gamma^5 \lambda^a s )( (\bar{s} \gamma_\nu \gamma^5 \lambda^a u ) \rangle_{\cal ST} -\frac{5}{6}m_N^3(A_4^u +A_4^s)\rho, \nonumber \\
\label{ope-medium}
\end{eqnarray}
where we have included the dimension 6 spin 2 operators, which vanishes in the vacuum.  Furthermore, while these operators have non-vanishing expectation values in general, we will take the medium expectation values to be zero based on the vacuum saturation hypothesis together with the chiral order parameter arguments given in the next section.  

$\Pi^{e(o)}$ denotes even(odd) dimensional terms of correlation function. $\langle {\cal O}\rangle_\rho$ is the condensate of operator ${\cal O}$ in nuclear matter, where we use the linear density approximation: $\langle {\cal O}\rangle_\rho=\langle {\cal O}\rangle_0+\langle {\cal O}\rangle_N  \rho$ with $\rho$ being the baryon density.  $\langle\bar{u}u\rangle_N$, $\langle\bar{s}s\rangle_N$, and $\langle G^2\rangle_N$ are the nucleon matrix elements.  $A_n^q(\mu^2)=2\int_0^1
x^{n-1}\{q(x,\mu^2)+(-1)^n \bar{q}(x,\mu^2)\}dx$, where $q(x,\mu^2)$ and $\bar{q}(x,\mu^2)$ are, respectively,
quark and antiquark distribution functions in the nucleon at scale $\mu^2$, and are defined through the twist-two operators $\langle {\cal ST}(\bar{q}\gamma_{\mu_1} D_{\mu_2} ... D_{\mu_n}q(\mu^2))\rangle_N  
=(-i)^{n-1}A_n^q(\mu^2)\frac{T_{\mu_1 ...\mu_n}}{2m_N},$ where ${\cal ST}$ means `symmetric and traceless',  and expressed on the right side  with the tensor $T_{\mu_1 \cdots\mu_n}$.

\subsubsection{Order parameter}

The difference in the correlation function between the vector and axial vector current has the following form up to dimension 6 operators
\begin{eqnarray}
\Pi^{AA}-\Pi^{VV} &=& \frac{2}{Q^2} m_s \langle\bar{u}u\rangle_0 -\frac{2 \pi}{Q^4} \alpha_s \bigg( \langle (\bar{u} \gamma_\mu \lambda^a s )( (\bar{s} \gamma_\mu  \lambda^a u ) \rangle - \langle (\bar{u} \gamma_\mu \gamma^5 \lambda^a s )( (\bar{s} \gamma_\mu \gamma^5 \lambda^a u ) \rangle
 \bigg) \nonumber \\
 && + \frac{8 \pi}{3Q^6}q^\mu q^\nu \alpha_s \bigg( \langle (\bar{u} \gamma_\mu \lambda^a s )( (\bar{s} \gamma_\nu  \lambda^a u ) \rangle_{\cal ST} - \langle (\bar{u} \gamma_\mu \gamma^5 \lambda^a s )( (\bar{s} \gamma_\nu \gamma^5 \lambda^a u ) \rangle_{\cal ST}
 \bigg). \label{order}
\end{eqnarray}
The second line of Eq.~(\ref{order}) corresponds to the twist-4 matrix element given in Ref.\cite{Choi:1993cu,Lee:1993ww,Friman:1999wu}.  Since the difference in Eq.~(\ref{order}) is a chiral order parameter, the operators at each dimension are proportional to the chiral symmetry breaking effects; $m_s \langle \bar{u} u \rangle$ at dimension 4 and $\langle \bar{s} s \rangle \langle \bar{u} u \rangle$ at dimension 6 after vacuum saturation.  This is so because the difference are proportional to the chiral symmetry breaking in the strangeness sector multiplied by those in the light quark sector.  Therefore, 
vacuum saturation hypothesis should be a good approximation for these combination of 4-quark operators, for which the twist-4 matrix do not contribute.  Therefore, even in the medium, one can estimate the change of the combination of 4-quark operators by just using the vacuum saturation hypothesis and change the light and strange quark condensate as is typically done in the  factorization formula of the 4-quark condensate in medium.

\subsubsection{Results}

In a numerical analysis recently performed in Ref.~\cite{Song:2018plu}, we find that the upper limit of the mass shift of $K_1^-$ ($K_1^+$) in nuclear matter is as large as -249 (-35) MeV.

\section{Phenomenological observations}

The dominant hadronic decay channels are given in Table II.
This means that with the kaon beam, the $K^*$ ($K_1$) can be produced by the $\pi$ ($\rho$) exchange with the nucleon. 
\begin{table}[tbh]
\caption{Dominant hadronic decay channels of $K^*$ and $K_1$ meson.}
\label{t1}
\begin{tabular}{||c|c|c||c|c|c||}
\hline
$J^{PC}=1^{-}$ & Decay Mode & Fraction & $J^{PC}=1^{+}$ & Decay mode & Fraction \\
\hline
$K^*(892)$ & $K \pi$ & 100\% & $K_1(1270)$ & $K \rho$ &  42\% \\
 &  &  &  & $K^* \pi$ &  16\% \\
\hline
\end{tabular}
\end{table}

As discussed before, the chiral partnership between the $K^*$ and $K_1$ exists between the same charge states.  If a $K^-$ beam is used to produce these particles,  the expected final states are given for $K_1$
\begin{eqnarray}
K_1^- \rightarrow \bigg(
\begin{array}{c}
\rho^0 K^-  \\
\rho^- \bar{K}^0   
\end{array} 
 \bigg(
\begin{array}{c}
\pi^0 K^{*-}  \\
\pi^- \bar{K}^{* 0}  
\end{array} 
, ~~~~
\bar{K}_1^0 \rightarrow \bigg(
\begin{array}{c}
\rho^+ K^-  \\
\rho^0 \bar{K}^0   
\end{array} 
 \bigg(
\begin{array}{c}
\pi^+ K^{*-}  \\
\pi^0 \bar{K}^{* 0}  
\end{array} ,
\nonumber
\end{eqnarray}
and for $K^*$ 
\begin{eqnarray}
K^{*-} \rightarrow \bigg(
\begin{array}{c}
\pi^0 K^{-}  \\
\pi^- \bar{K}^{ 0}  
\end{array} 
, ~~~~
\bar{K}^{*0} \rightarrow \bigg(
\begin{array}{c}
\pi^+ K^{-}  \\
\pi^0 \bar{K}^{ 0}  
\end{array} ,
\nonumber
\end{eqnarray}

\section{Conclusion}

After decades-long  attempts  to measure the mass shift and understand the origin of hadron mass,  it became clear that one has to analyze hadrons with small vacuum width.  Also, to identify the effect of chiral symmetry breaking effects, one has to start by looking at chiral partners.  Such consideration inevitably led us to consider $K^*$ and $K_1$ in matter\cite{Song:2018plu}.  In particular, with the kaon beam at JPARC, the possibility of observing both of these particles in a nuclear target experiment is quite feasible.  The problem of final state interaction from hadronic decays can be overcome by combining the study of excitation energy\cite{Metag:2017yuh}.  Once the masses and mass difference of $K^*$ and $K_1$ mesons are measured, one will be a closer to understanding the origin of the hadron masses and the effects of chiral symmetry breaking in them. 

\section{Acknowledgements}

This work was supported by Samsung Science and Technology Foundation under Project Number SSTF-BA1901-04.  
The author thanks Tetsuo Hatsuda and Taesoo Song for the collaboration on $K_1$ sum rule\cite{Song:2018plu}, which comprises a main theme of this talk.


\begin{thebibliography}{9}
\bibitem{Song:2018plu} 
  T.~Song, T.~Hatsuda and S.~H.~Lee,
  Phys.\ Lett.\ B {\bf 792}, 160 (2019).

\bibitem{Hatsuda:1985eb}
  T.~Hatsuda and T.~Kunihiro,
  Phys.\ Rev.\ Lett.\  {\bf 55}, 158 (1985).

\bibitem{Brown:1991kk}
  G.~E.~Brown and M.~Rho,
  Phys.\ Rev.\ Lett.\  {\bf 66}, 2720 (1991).

\bibitem{Hatsuda:1991ez}
  T.~Hatsuda and S.~H.~Lee,
  Phys.\ Rev.\ C {\bf 46}, no. 1, R34 (1992).

\bibitem{Klingl:1997kf}
  F.~Klingl, N.~Kaiser and W.~Weise,
  Nucl.\ Phys.\ A {\bf 624}, 527 (1997).
%
  R.~Rapp and J.~Wambach,
  Adv.\ Nucl.\ Phys.\  {\bf 25}, 1 (2000).
%
  S.~Leupold, V.~Metag and U.~Mosel,
  Int.\ J.\ Mod.\ Phys.\ E {\bf 19}, 147 (2010).
%
  P.~Gubler and W.~Weise,
  Nucl.\ Phys.\ A {\bf 954}, 125 (2016).



\bibitem{Hayano:2008vn}
 R.~S.~Hayano and T.~Hatsuda,
  Rev.\ Mod.\ Phys.\  {\bf 82}, 2949 (2010).



\bibitem{Naruki:2005kd}
  M.~Naruki {\it et al.},
  Phys.\ Rev.\ Lett.\  {\bf 96}, 092301 (2006).


\bibitem{Muto:2005za}
  R.~Muto {\it et al.}  [KEK-PS-E325 Collaboration],
  Phys.\ Rev.\ Lett.\  {\bf 98}, 042501 (2007).

\bibitem{Ichikawa:2018woh}
  M.~Ichikawa {\it et al.},
  arXiv:1806.10671 [physics.ins-det].

\bibitem{Trnka:2005ey}
  D.~Trnka {\it et al.}  [CBELSA/TAPS Collaboration],
  Phys.\ Rev.\ Lett.\  {\bf 94}, 192303 (2005).

\bibitem{Gubler:2016djf}
  P.~Gubler, T.~Kunihiro and S.~H.~Lee,
  Phys.\ Lett.\ B {\bf 767}, 336 (2017).

\bibitem{DuttMazumder:2000ys}
  A.~K.~Dutt-Mazumder, R.~Hofmann and M.~Pospelov,
  Phys.\ Rev.\ C {\bf 63}, 015204 (2001).


\bibitem{Thomas:2005dc}
  R.~Thomas, S.~Zschocke and B.~Kampfer,
  Phys.\ Rev.\ Lett.\  {\bf 95}, 232301 (2005).


\bibitem{Leupold:2009kz}
  S.~Leupold, V.~Metag and U.~Mosel,
  Int.\ J.\ Mod.\ Phys.\ E {\bf 19}, 147 (2010).

\bibitem{Metag:2017yuh}
  V.~Metag, M.~Nanova and E.~Y.~Paryev,
  Prog.\ Part.\ Nucl.\ Phys.\  {\bf 97}, 199 (2017).



\bibitem{Dickson:2016gwc}
  R.~Dickson {\it et al.} [CLAS Collaboration],
  Phys.\ Rev.\ C {\bf 93}, 065202 (2016).

\bibitem{Lee:2013es}
  S.~H.~Lee and S.~Cho,
  Int.\ J.\ Mod.\ Phys.\ E {\bf 22}, 1330008 (2013).




\bibitem{Suzuki:1993yc}
  M.~Suzuki,
  Phys.\ Rev.\ D {\bf 47}, 1252 (1993).

\bibitem{Hatsuda:1997ev}
  T.~Hatsuda,
  arXiv:nucl-th/9702002.

\bibitem{Manousakis:1986jh} 
  E.~Manousakis and J.~Polonyi,
  Phys.\ Rev.\ Lett.\  {\bf 58}, 847 (1987).

\bibitem{Shifman:1980ui} 
  M.~A.~Shifman,
  Nucl.\ Phys.\ B {\bf 173}, 13 (1980).

\bibitem{Dosch:1988ha} 
  H.~G.~Dosch and Y.~A.~Simonov,
  Phys.\ Lett.\ B {\bf 205}, 339 (1988).
  
\bibitem{Lee:1989qj} 
  S.~H.~Lee,
  Phys.\ Rev.\ D {\bf 40}, 2484 (1989).
  
\bibitem{Lee:2008xp} 
  S.~H.~Lee and K.~Morita,
  Phys.\ Rev.\ D {\bf 79}, 011501 (2009).

\bibitem{BC} T. Banks and A. Casher, Nucl. Phys.  {\bf B 169} 103 (1980).

\bibitem{Cohen:1996ng}
  T.~D.~Cohen,
  Phys.\ Rev.\ D {\bf 54}, 1867 (1996).

\bibitem{Lee:1996zy}
  S.~H.~Lee and T.~Hatsuda,
  Phys.\ Rev.\  D {\bf 54}, 1871 (1996).

\bibitem{tHooft:1986ooh} 
  G.~'t Hooft,
  Phys.\ Rept.\  {\bf 142}, 357 (1986).


 
\bibitem{Choi:1993cu} 
  S.~Choi, T.~Hatsuda, Y.~Koike and S.~H.~Lee,
  Phys.\ Lett.\ B {\bf 312}, 351 (1993).
  
\bibitem{Lee:1993ww} 
  S.~H.~Lee,
  Phys.\ Rev.\ D {\bf 49}, 2242 (1994).




\bibitem{Friman:1999wu}
  B.~Friman, S.~H.~Lee and H.~C.~Kim,
  Nucl.\ Phys.\ A {\bf 653}, 91 (1999).


\end{thebibliography}
\end{document}